\documentclass[prc,href,aps,a4paper,groupedaddress,superscriptaddress,nofootinbib,showpacs,twocolumn]{revtex4}

\usepackage{hyperref}
\usepackage{graphicx}
\usepackage{amsfonts}
\usepackage{amssymb}
\usepackage{amsmath}
\usepackage{natbib}
\usepackage{dcolumn}
\usepackage{bm} 
\usepackage{multirow}
\usepackage{epsfig,ulem}
\usepackage{amsfonts}
\usepackage{hhline,color}
\usepackage{pifont}
\usepackage{lmodern}
\usepackage[utf8]{inputenc}

\newcommand{\bwt}{\begin{widetext}}
\newcommand{\ewt}{\end{widetext}}
\newcommand{\beq}{\begin{equation}}
\newcommand{\eeq}{\end{equation}}
\newcommand{\bea}{\begin{eqnarray}}
\newcommand{\eea}{\end{eqnarray}}
\begin{document}
\title{ Warm unstable asymmetric nuclear matter: critical properties and the density
dependence of the symmetry energy}

\author{N. Alam}
\email {naosad.alam@saha.ac.in}
\affiliation{Saha Institute of Nuclear Physics, HBNI, 1/AF Bidhannagar, Kolkata 700064, India.}
\author{H. Pais}
\affiliation{CFisUC, Department of Physics, University of Coimbra, 3004-516 Coimbra, Portugal.} 
\author{C. Provid{\^e}ncia}
\affiliation{CFisUC, Department of Physics, University of Coimbra, 3004-516 Coimbra, Portugal.} 
\author{B. K. Agrawal}
\affiliation{Saha Institute of Nuclear Physics, HBNI, 1/AF Bidhannagar, Kolkata 700064, India.}

\date{\today}    
\begin{abstract}

The spinodal instabilities in hot asymmetric nuclear matter  and some
important critical  parameters derived thereof are studied using
six different families of  relativistic mean-field (RMF) models.
The slopes of the symmetry energy coefficient vary over a wide range
within each family. The critical densities and proton
fractions are more sensitive to the symmetry energy slope parameter at
temperatures much below its critical value ($T_c\sim$14-16 MeV).
The  spread in the critical proton fraction at a given symmetry energy
slope parameter is noticeably larger   near $T_c$, indicating that the
warm equation of state of asymmetric nuclear matter at sub-saturation densities is
not sufficiently constrained. The distillation effects are sensitive
to the density dependence of the symmetry energy at low
temperatures which tend to wash out with increasing temperature.
 
\end{abstract}


\maketitle

\section{Introduction}

Core-collapse supernovae (CCSN) \cite{CCSN} are one of the most energetic
events in the Universe. Matter can reach temperatures up to $\sim 20$
MeV and the density at bounce of the collapsing core goes up to 1.5 -
2.0 times the nuclear saturation density.  During the collapse,
matter does not have enough time to reach $\beta$-equilibrium conditions
\cite{Stone07}, because the event timescale  is believed to be of the
order of seconds, and usually a fixed proton fraction of $y_p \sim 0.3$
\cite{Mezzacappa05} is considered for the calculation of the EoS. The
reader can refer to \cite{oertel16} and references therein for a recent
review on the relevant thermodynamics and composition for the equation
of state for CCSN, compact stars and  compact stars mergers.
At densities below nuclear saturation, light \cite{clusters} and heavy
clusters \cite{pasta} can form,  and they can modify the neutrino
transport, which will affect the cooling of the proto-neutron star
\cite{coolingPNS}, as  neutrinos play a considerable role in the
development of the shock wave during the collapse  \cite{shock}. The
determination of the region of densities, proton fractions and
temperatures where these instabilities exist is, therefore, very
important for core-collapse simulations.

Critical properties of hot asymmetric and symmetric nuclear
matter may be studied with heavy ion collisions, in particular,
with nuclear reactions that involve the formation of compound
nuclei or multifragmentation. These data will be important to
constrain the CCSN EoS. As shown in Ref. \cite{Buyukcizmeci13},
the expected range of densities and temperatures for CCSN matter
just before bounce lie in  the typical ($\rho$,$T$) space, $\rho
\sim 0.05\rho_0 - 0.4\rho_0$ and $T \sim 3-8$ MeV, for nuclear
multifragmentation reactions.  In Ref. \cite{lourenco16},  a
compilation of the critical temperatures determined from experimental
data \cite{natowitz02,karnaukhov03,elliott13,experimental}, and which
generally fall above 16 MeV, is compared with  theoretically determined
ones from RMF models with non-linear sigma models that have an effective
mass at saturation in the range 0.58$\le m^*/M\le 0.65$, as obtained
from finite nuclei spin-orbit splittings, and the incompressibility in
the range 250$\le K_0\le 315$ MeV, as proposed in \cite{stone14}. Under
these conditions, it was shown that  the critical temperature from
RMF models satisfies $14.2\le T_c\le 16.1$ MeV, far from the value
proposed in \cite{elliott13}, where the authors have analysed  six sets
of experimental data, two involving the formation of compound nuclei
and four multifragmentation processes, and have determined  a critical
temperature of $T_c=17.9\pm 0.4$ MeV.  In order to be able to reproduce
the experimental critical temperature, and within RMF models that only
include $\sigma$ non-linear terms, the finite nuclei spin-orbit constraint
had to be relaxed in \cite{lourenco16} and a larger nuclear effective
mass chosen. However, these experimental constraints for the critical
temperature, above which matter is stable against clusterization, are
for symmetric matter. Constraints for asymmetric hot matter are missing.

In Refs. \cite{avanciniPasta}, the authors used several methods to
determine the crust-core transition, {including a Thomas Fermi calculation
of the inner crust and the thermodynamical and dynamical spinodals},
and they showed that for finite temperature and fixed proton fractions
(CCSN conditions), the thermodynamical method gave quite similar results
to more demanding calculations, like the Thomas-Fermi calculation.

The  thermodynamical spinodal, the boundary of the instability region
identified by a negative curvature free energy, is  determined by
equating the free energy curvature to zero. In Ref. \cite{avancini06},
the authors used the thermodynamical approach to analyse the liquid-gas
phase transition in  warm asymmetric nuclear matter, as well as stellar
matter, within RMF models with and without density-dependent couplings. In
particular, they discussed the isospin distillation effect, that is,
the different isospin content of each phase, with the gas being more
neutron-rich and the liquid phase with a proton fraction close to
symmetric matter. They showed that this effect is not so strong when
considering models with density-dependent couplings. They calculated
for each temperature the critical points of the spinodal, that is,
the two points where the pressure is maximum, together with the
critical temperature of the system, i.e., the temperature at which the
instability region melts. The two models with density dependent couplings
were shown to have a region of instabilities that extended to smaller
proton fractions and similar densities ranges when compared with models
with constant couplings, but no discussion was done on the connection
of these results with the density dependence of the symmetry energy.
In Ref. \cite{pais16}, the authors also used this method together with
other two to calculate the crust-core transition and pressure at zero
temperature, and using two of the families that are also going to be used
in this work. They observed that this calculation gives a good estimation
of the transition, like the authors of Ref. \cite{Ducoin11} also found.

 In this work, the critical parameters for hot asymmetric nuclear matter
for six different families of RMF models are studied using the
thermodynamical method.  These six families of models have been
built from three different appropriately calibrated base models. An
extra term that couples the $\rho$-meson either to the $\sigma$
or $\omega$-meson is added to each of the base models to yield wide
variations in the symmetry energy slope $L$.  The effect of $L$ on the
critical temperature, density and proton fraction is then explored. We
also compare our findings with experimental results from references
\cite{natowitz02,karnaukhov03,elliott13,experimental} for  the critical
temperature,  and the theoretical study of Ref. \cite{lourenco16}.

\section{Formalism}

We give a brief summary of the  RMF formalism in the
first subsection, and of the thermodynamical spinodal calculation and respective critical
points  in the second subsection.

\subsection{Extended RMF Lagrangian}

We consider a set of families, each one characterized by the same
isoscalar properties, which are described by the  scalar-isoscalar field
$\phi$ with mass $m_s$, associated to the $\sigma$-meson, and  the
vector-isoscalar field $V^{\mu}$ with mass $m_v$ associated to the
$\omega$-meson. The members of each family differ by their isovector properties which will be
determined by the vector-isovector field $\mathbf b^{\mu}$  with mass
$m_\rho$, associated to the $\rho$-meson,  and the  non-linear terms that couple the $\rho$-meson to the
$\sigma$ and/or the $\omega$-mesons. Nucleons, with mass $M$ and
described by the spinors $\psi_i $,  interact with and through the
$\sigma$, $\omega$ and $\rho$-mesons, 
according to the Lagrangian density:
$$
{\cal L}=\sum_{i=p,n} {\cal L}_i + {\cal L}_{\sigma} + {\cal
  L}_{\omega}  + {\cal L}_{\rho} + {\cal L}_{\sigma\omega\rho} \, ,
$$
where the nucleon Lagrangian reads
$$
{\cal L}_i=\bar \psi_i\left[\gamma_\mu i D^{\mu}-M^*\right]\psi_i \, ,
$$
with
$$
i D^{\mu}=i\partial^{\mu}-g_v V^{\mu}-
\frac{g_{\rho}}{2}  {\boldsymbol\tau} \cdot \mathbf{b}^\mu - e A^{\mu}
\frac{1+\tau_3}{2} \,,
$$
$$
M^*=M-g_s \phi  \, .
$$
The mesonic Lagrangians are:
\begin{eqnarray}
{\cal L}_\sigma&=&+\frac{1}{2}\left(\partial_{\mu}\phi\partial^{\mu}\phi
-m_s^2 \phi^2 - \frac{1}{3}\kappa \phi^3 -\frac{1}{12}\lambda\phi^4\right),\nonumber\\
{\cal L}_\omega&=&-\frac{1}{4}\Omega_{\mu\nu}\Omega^{\mu\nu}+\frac{1}{2}
m_v^2 V_{\mu}V^{\mu} + \frac{1}{4!}\xi g_v^4 (V_{\mu}V^{\mu})^2, \nonumber \\
{\cal L}_\rho&=&-\frac{1}{4}\mathbf B_{\mu\nu}\cdot\mathbf B^{\mu\nu}+\frac{1}{2}
m_\rho^2 \mathbf b_{\mu}\cdot \mathbf b^{\mu}, \nonumber\\
\end{eqnarray}
where
$\Omega_{\mu\nu}=\partial_{\mu}V_{\nu}-\partial_{\nu}V_{\mu} ,
\quad \mathbf B_{\mu\nu}=\partial_{\mu}\mathbf b_{\nu}-\partial_{\nu} \mathbf b_{\mu}
- g_\rho (\mathbf b_\mu \times \mathbf b_\nu)$, and $\boldsymbol \tau$ are the Pauli matrices.
The mesonic Lagrangian is supplemented with the following non-linear terms that mix  the $\sigma, \omega$, and $\mathbf{\rho}$ mesons  up to quartic order \cite{quarticA,quarticB,quartic1,quartic2}, 
 \begin{eqnarray}
\label{eq:lnon-lin}
{\cal L_{\sigma\omega\rho}} & =&\Lambda_{1\sigma}g_{s}g_{\rho }^{2}\phi \mathbf b_{\mu}\cdot \mathbf b^{\mu} 
+\Lambda_{\sigma} g_s^2 g_\rho^2 \phi^2 \mathbf b_{\mu}\cdot \mathbf b^{\mu} \nonumber \\
&+&\Lambda_{v} g_v^2 g_\rho^2 \mathbf b_{\mu}\cdot \mathbf b^{\mu}\, V_{\mu}V^{\mu} .
\end{eqnarray}
The parameters of RMF  models, which in the present case are the couplings
$g_s$, $g_v$, and $g_{\rho}$ of the mesons
to the nucleons, the nucleon bare mass $M$, the meson masses, the self-interacting coupling
constants, $\kappa$, $\lambda$, and $\xi$, and the coupling constants
of the non-linear mixing terms, $\Lambda_v, \Lambda_\sigma, \Lambda_{1\sigma}$,
 are fixed to nuclear properties obtained
experimentally, and to astrophysical constraints \cite{quartic1,quartic2}.

The free energy density is obtained from the relation
\begin{equation}
{\cal{F}}={\cal{E}}- T\cal{S},
\end{equation}
with the energy density $\cal{E}$ given by
\begin{eqnarray}
\cal{E}&=&
\sum_{i=p,n} E_i+g_v V_0 \rho_v  + g_\rho b_0  \rho_3  \nonumber\\
 &+&\frac{1}{2} [(\nabla \phi_0 )^2 +  m_s^2\phi_0^2 ]   
+\frac{\kappa}{3!} \phi_0^3  +\frac{\lambda}{4!} \phi_0^4  \nonumber\\
&-&\frac{1}{2}\left[(\nabla V_0 )^2 +m_v^2 V_0^2 
+\frac{\xi g_v^4}{12} V_0^4 \right]
 -\frac{1}{2} \left[(\nabla b_0  )^2 
+m_{\rho}^2 b_0^2 \right]\nonumber\\
&-&\left(\Lambda_v\,g_v^2\,V_0^2 -\Lambda_\sigma\,g_s^2\,\phi^2 -\Lambda_{1\sigma}\,g_s\,\phi\right) \,g_\rho^2\, b_0^2 ,
\label{enfun} 
\end{eqnarray}
where the energies $E_i$ are
\begin{eqnarray}
E_i=&&\frac{1}{\pi^2}\int dp\, p^2\, \epsilon_i^*
  \left(f_{i+}+f_{i-}\right) \, ,\, i=p,n,
\end{eqnarray}
with  the  equilibrium distribution functions  defined as 
\begin{eqnarray}
f_{i\pm}&=&\frac{1}{1+\exp
\left[{(\epsilon_i^*\mp \nu_i)/T}\right] } \, , 
\end{eqnarray}
 $\epsilon_i^*=\sqrt{p^2+{M_i^*}^2}$,
$M_i^*  = M- {g_{s}}\phi$, and
the nucleons effective chemical potential
\begin{equation}
\nu_i=\mu_i - g_{v} V_0 - {g_{\rho}}~  t_{3 i}~ b_0,
\end{equation}
where $t_{3i}$ is the third component of the isospin operator.
The entropy density $\cal{S}$ is calculated considering the nucleons as quasiparticles
\begin{eqnarray}
\cal{S}&=&  -\sum_{i=n,p}\int \frac{d^3 p}{4\pi^3} ~
\left[ f_{i+}  \ln f_{i+}   +\left(1- f_{i+} \right) \ln
            \left(1-f_{i+}  \right)\right. \nonumber\\
&+&
\left.( f_{i+} \leftrightarrow f_{i-} )
\right].
\end{eqnarray}

\subsection{Stability Conditions}

In the present study, we determine the region of instability of
nuclear matter constituted by protons and neutrons by calculating the
spinodal surface in the $(\rho_p, \, \rho_n,\, T)$ space. Stability
conditions for asymmetric matter  impose that the curvature matrix of
the free energy density \cite{bctl98,ms,avancini06}
\begin{equation}
{\cal C}_{ij}=\left(\frac{\partial^2{\cal F}}{\partial \rho_i\partial\rho_j}
\right)_T,
\label{stability}
\end{equation}
 is positive.
Eq. (\ref{stability}) can be rewritten in the form \begin{equation}
\mathcal{C}=\left( 
\begin{array}{ll}
\frac{\partial \mu _{n}}{\partial \rho _{n}} & \frac{\partial \mu _{n}}{
\partial \rho _{p}} \\ 
\frac{\partial \mu _{p}}{\partial \rho _{n}} & \frac{\partial \mu _{p}}{
\partial \rho _{p}}
\end{array}
\right) ,  \label{stability1}
\end{equation}
imposing 
\begin{equation}
\mbox{Tr}(\mathcal{C})>0,  \label{stab1}
\end{equation}
\begin{equation}
\mbox{Det}(\mathcal{C})>0,  \label{stab2}
\end{equation}
to fulfil the stability conditions.
This is equivalent to requiring that the two eigenvalues
\begin{equation}
\lambda _{\pm }=\frac{1}{2}\left( \mbox{Tr}(\mathcal{C})\pm \sqrt{\mbox{Tr}(
\mathcal{C})^{2}-4\mbox{Det}(\mathcal{C})}\right) ,
\end{equation}
are positive. The largest eigenvalue is always positive and  the
instability region is delimited by the surface  $\lambda_-=0$. Interesting information
is given by the associated eigenvectors $\boldsymbol{\delta\rho^\pm}$, defined as
$$
\frac{\delta \rho ^{\pm }}{\delta \rho _{n}^{\pm }}=\frac{\lambda^{\pm }-
\frac{\partial \mu _{n}}{\partial \rho _{n}}}{\frac{\partial \mu _{n}}{
\partial \rho _{p}}}.
$$
In particular, the eigenvector associated with the eigenvalue that defines the
spinodal surface determines the instability direction, i.e. the direction
along which the free energy decreases. We will also calculate the
critical points for each temperature $T$, which are important to define
under which conditions the system is expected to be clusterized. These points satisfy
simultaneously  \cite{reid,avancini06}
\begin{eqnarray}
\mbox{Det}(\mathcal{C})&=&0\\
\mbox{Det}(\mathcal{M})&=&0,
\end{eqnarray}
with
\begin{equation}
\mathcal{M}=\left( 
\begin{array}{ll}
 {\cal C}_{11} & {\cal C}_{12}\\
 \frac{\partial {|\cal C}|}{\partial \rho_p}& \frac{\partial {|\cal C}|}{\partial \rho_n}
\end{array}
\right) .  
\label{critical}\end{equation}
The thermodynamical spinodals and respective  critical points will be
calculated for a series of models in the next section.

\section{Results}

In this section, we first briefly describe the different families of the
RMF models used for the current study.  Next, we present our results for
the spinodal instabilities and critical points in hot asymmetric matter at different temperatures.
The effect of the symmetry energy slope parameter, $L$, on this quantities will be addressed as well.

\subsection{Models}

 \begin{figure*}[!htb]
\begin{tabular}{cc}
 \includegraphics[width=0.4\linewidth,angle=0]{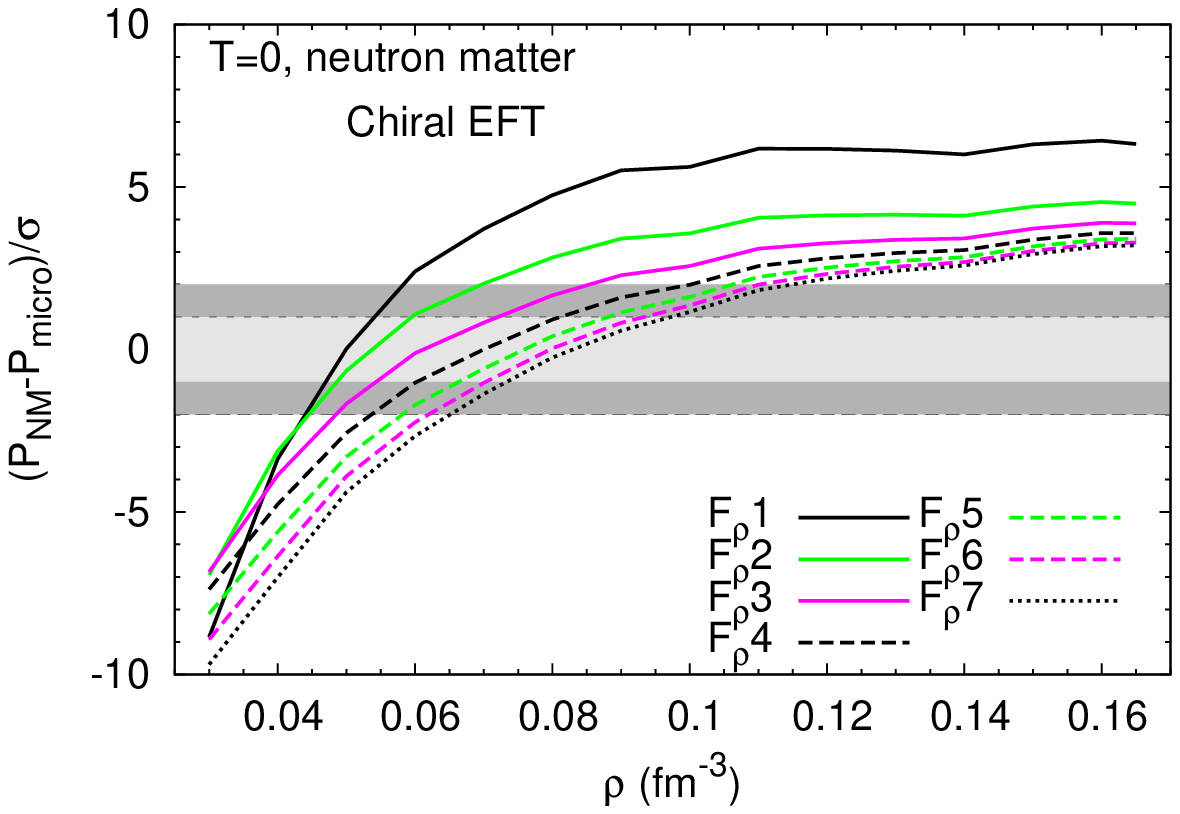}
&\includegraphics[width=0.4\linewidth,angle=0]{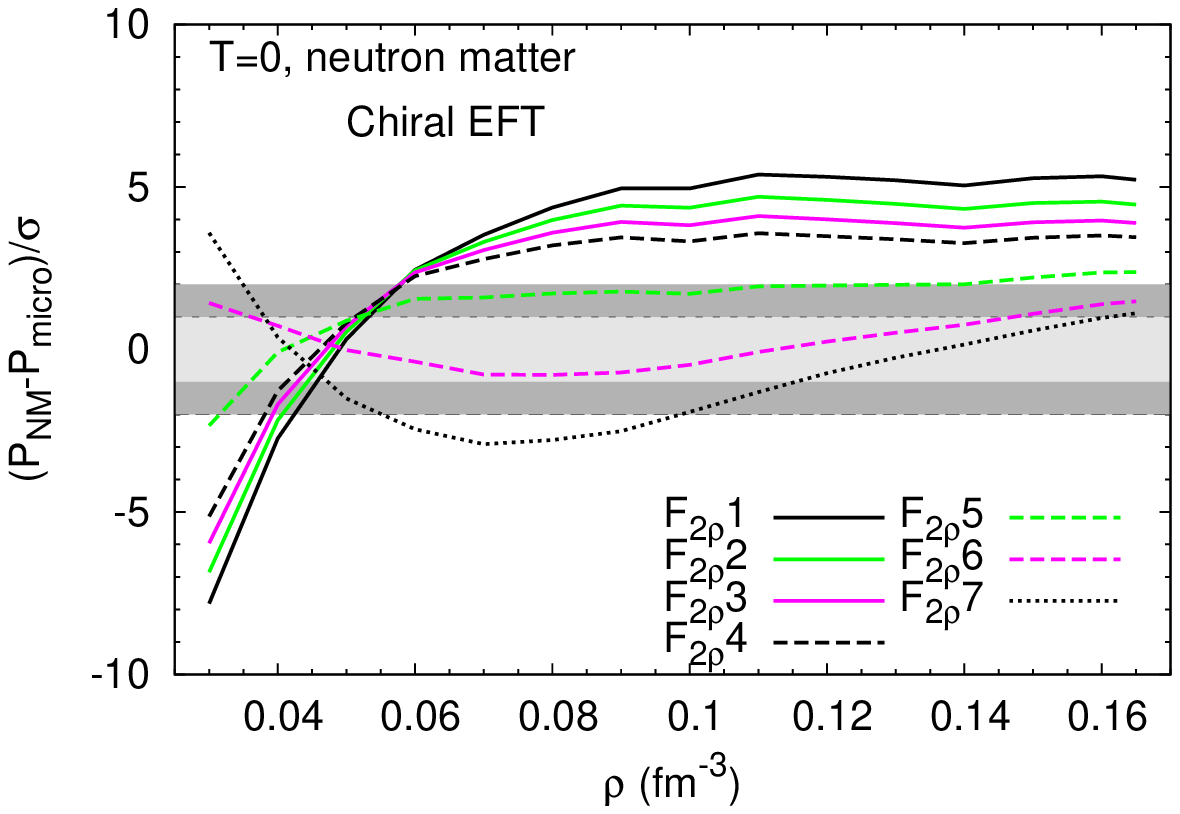}\\
 \includegraphics[width=0.4\linewidth,angle=0]{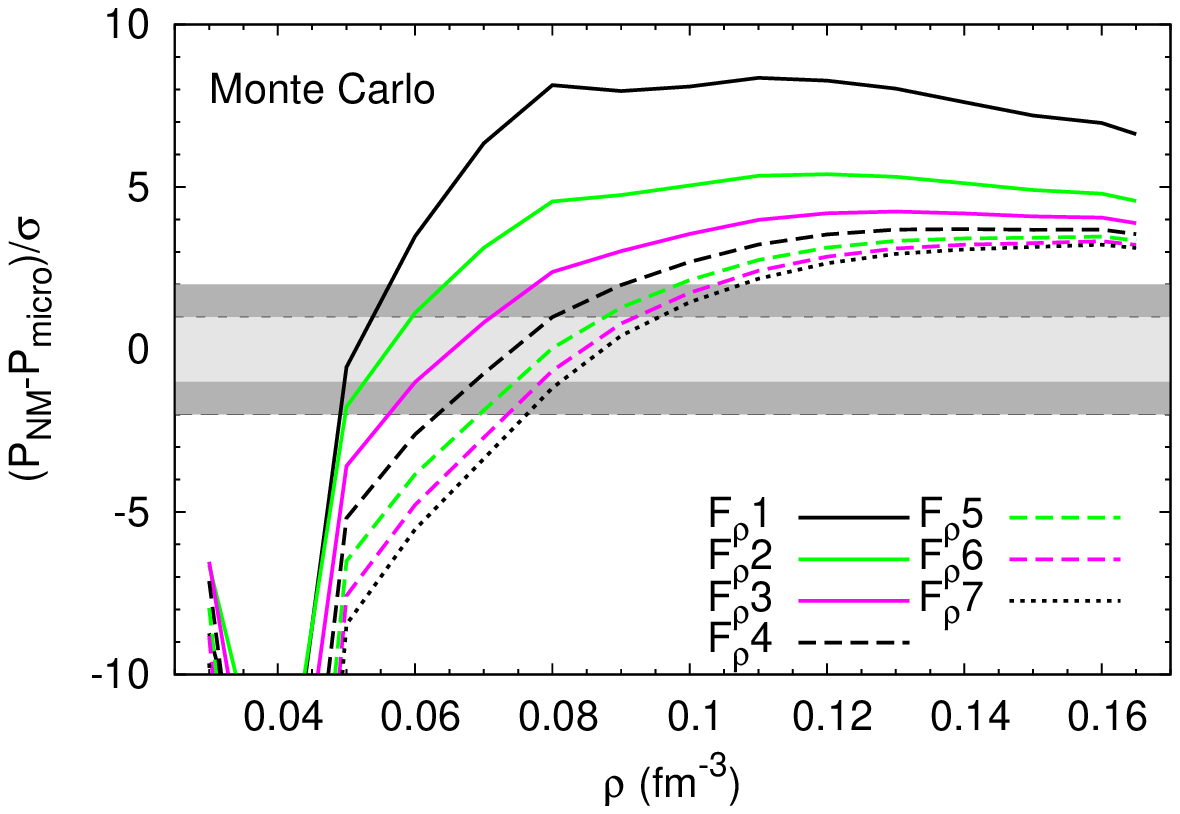}&
  \includegraphics[width=0.4\linewidth,angle=0]{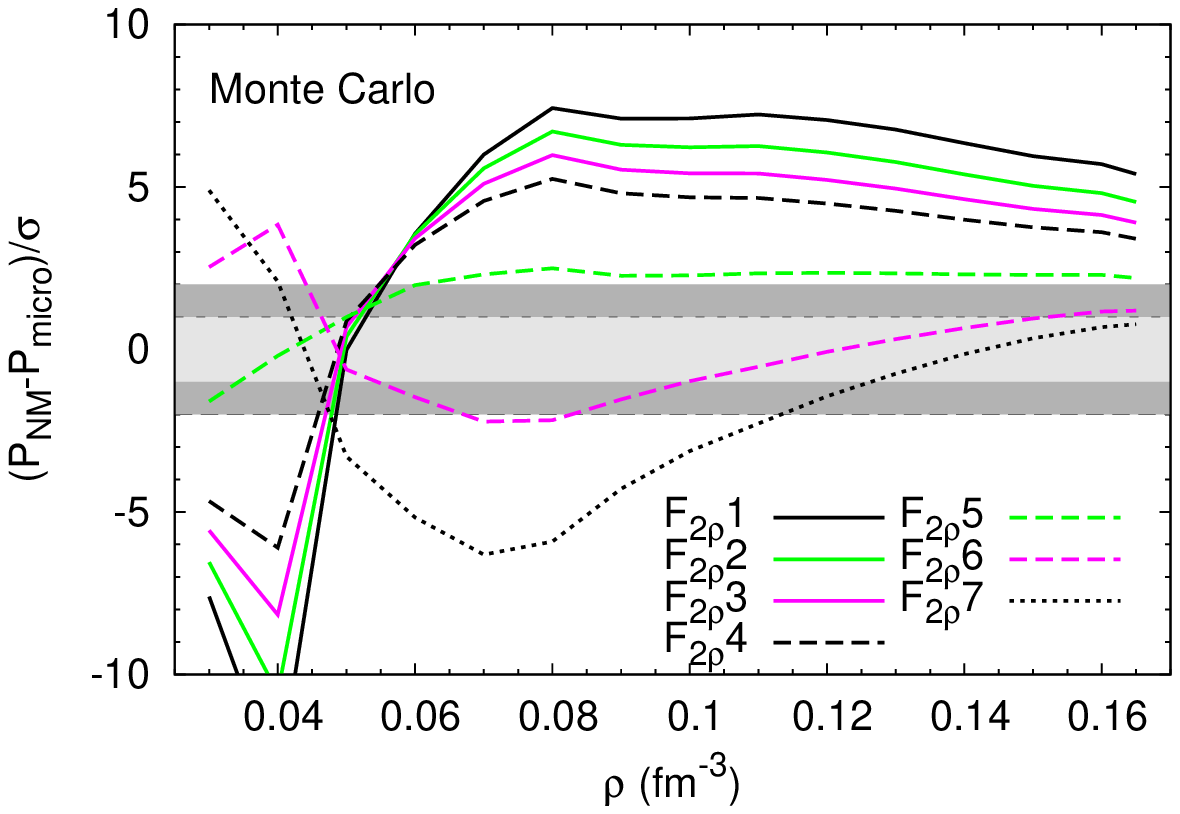}
\end{tabular}
 \caption{(Color online) Difference between the neutron matter pressure for 
 the F$_{\rho}$ (left panels) and F$_{2\rho}$ (right panels) families and the
 average pressure obtained from a chiral EFT \cite{Hebeler} (top) and Monte Carlo \cite{Gandolfi}
 (bottom) calculations, in units of the pressure uncertainty at each density, $\sigma=\Delta P$.
 The gray bands represent the calculation uncertainty (light) and twice this uncertainty (dark). }
 \label{fig1}
 \end{figure*}

In this work, we are going to consider six different families of RMF
models, namely, NL3$\omega\rho$ \cite{Carriere}, TM1$\omega\rho$
\cite{paisVlasov}, F$_{2\rho}$ \cite{frho,bka22}, NL3$\sigma\rho$
\cite{paisVlasov}, TM1$\sigma\rho$ \cite{paisVlasov}, and F$_{\rho}$
\cite{frho}. The NL3$\omega\rho$ and NL3$\sigma\rho$ (TM1$\omega\rho$ and
TM1$\sigma\rho$) families are obtained from the base model NL3 \cite{nl3}
(TM1\cite{tm1}). The F$_\rho$ and F$_{2\rho}$ families are obtained
from the base model BKA22 \cite{bka22}.  The families NL3$\omega\rho$,
TM1$\omega\rho$ and F$_{2\rho}$ include a quartic order cross-coupling
$\omega^2\rho^2$ term ($\Lambda_v\neq0$), whereas the NL3$\sigma\rho$ and
TM1$\sigma\rho$ families have a quartic order cross coupling $\sigma^2\rho^2$
term ($\Lambda_{\sigma}\neq0$). On the other hand,  a cubic
order cross-coupling $\sigma\rho^2$ term ($\Lambda_{1\sigma}\neq0$)  is included in the F$_{\rho}$ family.
The strengths of the  cross-couplings ($\Lambda_v,\Lambda_\sigma,$ and
$\Lambda_{1\sigma}$), and that of the coupling of the $\rho$-mesons to 
the nucleons ($g_\rho$), are appropriately adjusted to vary the slope
of symmetry energy over a wide range without compromising the
properties of the finite nuclei significantly.
The cross-couplings $\Lambda_v$ or $\Lambda_{\sigma}$ or
$\Lambda_{1\sigma}$ is increased (decreased) and accordingly $g_\rho$ is
also increased (decreased) in such a way that either the binding energy
of $^{208}$Pb nucleus is close the the experimental value or the symmetry
energy at density  $0.1$ fm$^{-3}$ is exactly the same as that for the base
model. Different combinations of the coupling strengths yield different 
behaviours for the density dependence of the symmetry energy.
The variants of NL3 and TM1 models
are obtained by varying $\Lambda_v$ or $\Lambda_\sigma$ and adjusting
$g_\rho$ in such a way that the symmetry energy at  $\rho=0.1$ fm$^{-3}$ 
is equal to the one obtained for the base models \cite{Horowitz,Carriere}.  The variants of BKA22
model (i.e. F$_{\rho}$ and F$_{2\rho}$ families) are obtained by varying
$\Lambda_v$ or $\Lambda_{1\sigma}$ and adjusting $g_\rho$ to reproduce
the binding energy of the $^{208}$Pb nucleus.  All the families of models
considered are consistent with the observational constraints imposed by the
measured mass ($\sim 2M_\odot$) of the pulsars J1614-2230 \cite{J1614} and
J0348+0432 \cite{J0348}, see e.g. \cite{Alam16} and references there in.
Besides these observational constrains, there are also experimental
results and first-principle calculations that can allow to set limits
on the stellar matter EoS. In Table \ref{tab1}, we present some bulk
properties of $^{208}$Pb nucleus as well as the neutron star maximum mass 
and corresponding radius obtained for the models with extreme
values of $L$ from each families.

In addition to these six families of models, we also consider as reference  two extra models
with density dependent couplings: DD2 \cite{dd2} and DDME2 \cite{ddme2}. In Ref. \cite{fortin16}, it
was shown that these two models satisfy a well accepted set of
laboratorial, theoretical and observational constraints. We
are, therefore, interested in comparing the behaviour of these models at finite
temperature with the behaviour of the six families of models we are going to analyse.

\begin{table}[t]
\caption{  \label{tab1}
The values of the binding energy per particle ($B/A$), charge radii ($r_c$ ), neutron radii ($r_n$) and
neutron skin thickness ($\Delta r_{\rm np}$) for $^{208}$Pb nucleus along with the maximum mass ($M_{max}$) of neutron
star and corresponding radius ($R_{max}$) obtained for some selected models.}
\begin{ruledtabular}
\vspace{0.5cm}
\begin{tabular}{ccccccc}
Model &  $B/A$ &  $r_c$  & $r_n$ & $\Delta r_{\rm np}$ & $M_{max} $& $R_{max}$ \\
      & (MeV)  & (fm)    & (fm)  &   (fm)              & (M$_\odot$)&  (km)  \\
    \hline
$F_\rho$1           & -7.871  & 5.529    & 5.751   & 0.280  & 1.99 & 11.77  \\
$F_\rho$7           & -7.871  & 5.559  &  5.680   & 0.179   & 1.97  & 11.33  \\
$F_{2\rho}$1        & -7.871  & 5.529    & 5.740   & 0.269 & 1.95  & 11.61 \\
$F_{2\rho}$7        & -7.870  & 5.555 &  5.649   & 0.152   & 1.93  & 11.06 \\
NL3                 & -7.878   & 5.518 & 5.740 &  0.280   & 2.78 & 13.29  \\
NL3${\sigma\rho}$6  & -7.913   & 5.535 & 5.662 &  0.185  & 2.77 & 13.14 \\
NL3${\omega\rho}$6  & -7.921   & 5.530 & 5.667 &  0.195  & 2.76 & 12.99    \\
TM1                 & -7.877   & 5.541 & 5.753 &  0.270  & 2.18  & 12.49  \\
TM1${\sigma\rho}$6  & -7.923   & 5.558 & 5.686 &  0.186  & 2.15 & 12.02  \\
TM1${\omega\rho}$6  & -7.791   & 5.552 & 5.689 &  0.195  & 2.13 & 11.97    \\
  \end{tabular}
  \end{ruledtabular}
\end{table}

\begin{figure}[!htb]
 \includegraphics[width=0.95\linewidth,angle=0]{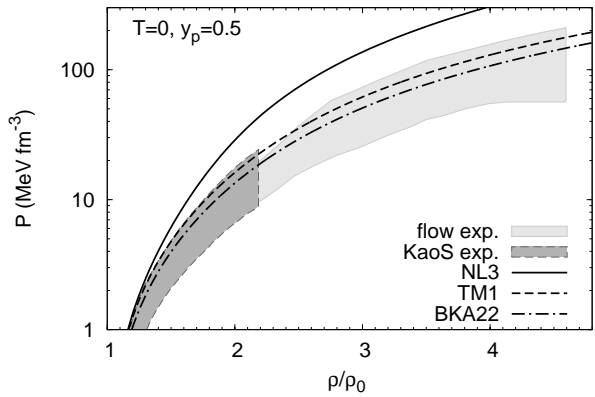} \\
 \caption{(Color online) Symmetric matter pressure as a function of the density for the NL3 (solid), TM1 (dashed)
 , and BKA22 (dash-dotted) models. The colored bands are the experimental results obtained from collective flow 
 data in heavy-ion collisions \cite{flow} (light gray) and from the KaoS experiment \cite{kaons} (dark gray). }
 \label{fig2}
 \end{figure}

 \begin{figure}[!htb]
 \includegraphics[width=0.95\linewidth,angle=0]{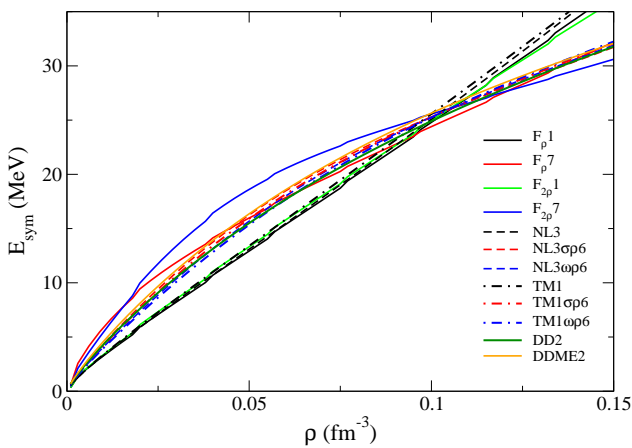} \\
 \caption{(Color online) Symmetry energy as a function of baryon density  for the F$_{\rho}$1, F$_{\rho}$7, F$_{2\rho}$1, F$_{2\rho}$7 (solid),
  NL3, NL3${\sigma\rho}$6, NL3${\omega\rho}$6 (dashed), 
 and TM1, TM1${\sigma\rho}$6, TM1${\omega\rho}$6 (dash-dotted) models.
 The DD2 (green) and DDME2 (orange) models are also represented for comparison.}
 \label{fig3}
 \end{figure}

In Fig.~\ref{fig1}, we compare the neutron matter pressure of the F$_\rho$ and $F_{2\rho}$ families with microscopic calculations
based on nuclear interactions derived from chiral effective field theory (EFT) \cite{Hebeler}, and quantum Monte Carlo techniques  with
realistic two- and three-nucleon interactions \cite{Gandolfi}. We show the difference from the neutron matter pressure of each model to the
microscopic results, normalized to the pressure uncertainty of the microscopic calculations, $\sigma=\Delta P$, at each density.  These
uncertainties are represented by light gray bands, and they indicate that the points that lie inside those bands are within the data
limits. Also shown are dark gray bands that denote twice the calculation uncertainties, $2\sigma$. 
 We observe that only F$_{2\rho}5$ and F$_{2\rho}$6 lie in the bands' limits. All the other models fail to satisfy these constrains.
Similarly, for other families, it was shown in Ref. \cite{paisVlasov} that only 4 models, NL3$\omega\rho6$, NL3$\sigma\rho6$, 
TM1$\omega\rho6$, and TM1$\sigma\rho6$, passed these microscopic constrains.

 In Fig.~\ref{fig2}, we show the EoSs for symmetric nuclear matter
 for the three base models considered, together with the
experimental results from collective flow data in heavy-ion collisions
\cite{flow}, and from the KaoS experiment \cite{kaons}.  The models of the NL3
family do not satisfy these constraints but  the EoSs for
the other models lie within the experimental bounds. However, the
modelling of flow in transport simulations is a complex process and,
therefore, these constraints should be considered with
care. Consequently,  we will also include the models of the NL3 family in
the present study.

We will be discussing the effect of the density dependence of 
symmetry energy on the extension of the  instability. To facilitate our discussions,
we show in Fig. \ref{fig3} the behaviour of symmetry energy
 at sub-saturation densities for the models with extreme values
of the slope $L$. The models with the largest $L$
have all a very similar behavior, showing an almost linear increase of
the symmetry energy with the density, typical of models that do not
have non-linear terms involving the $\rho$-meson. With respect to the
models with the smallest $L$,  the NL3$x6$ and
TM1$x6$ models have a similar behavior and   $L\sim
55$ MeV, showing a larger symmetry
energy below $\rho=0.1$ fm$^{-3}$ than the models with large
$L$. $F_{2\rho}7$ has a more extreme behavior due to its lower $L$,
$L=45$ MeV. The symmetry energy curves for the models corresponding 
to extreme values of $L$ cross each other at $\rho\sim 0.1$ fm$^{-3}$,
except for the $F_\rho$ family. 
The $F_\rho7$ crosses  $F_\rho1$ at a smaller density.

\subsection{Spinodal sections and critical points}

\begin{figure}[htb]
 \centering
 \includegraphics[width=1\linewidth,angle=0]{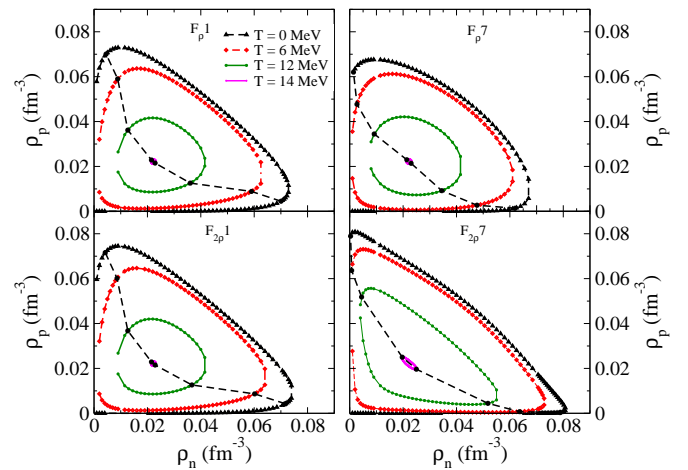}
 \caption{(Color online) Spinodal sections on the  ($\rho_n$, $\rho_p$)
   plane for $F_{\rho}$1 (top left), $F_{\rho}$7 (top right), $F_{2\rho}$1 (bottom left) and $F_{2\rho}$7 (bottom right) models at $T = 0, 6, 12 $
and $ 14 \hbox{ MeV}$.}
 \label{fig4}
 \end{figure}

 \begin{figure}[htb]
 \centering
\includegraphics[width=1\linewidth,angle=0]{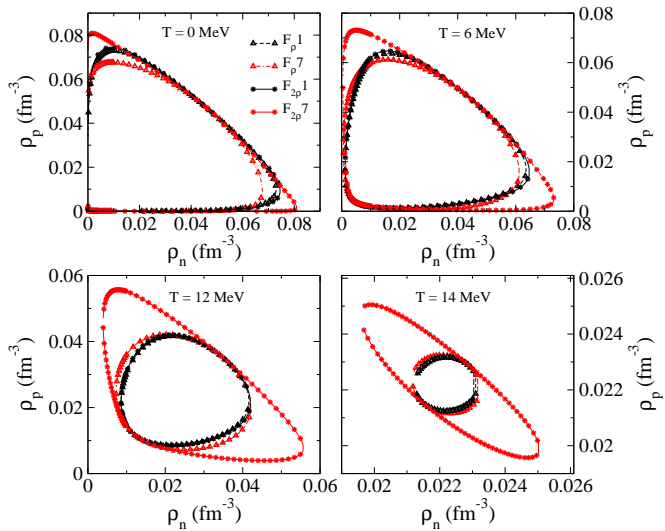} 
 \caption{(Color online) Spinodal sections on the  ($\rho_n$, $\rho_p$)
   plane for $F_{\rho}$1, $F_{\rho}$7, $F_{2\rho}$1 and
   $F_{2\rho}$7 models at $T = 0$ (top left), 6 (top right), 12 (bottom left) and 14 (bottom right)  MeV.}
 \label{fig5}
 \end{figure}

\begin{figure*}[htb]
\includegraphics[width=.9\linewidth,angle=0]{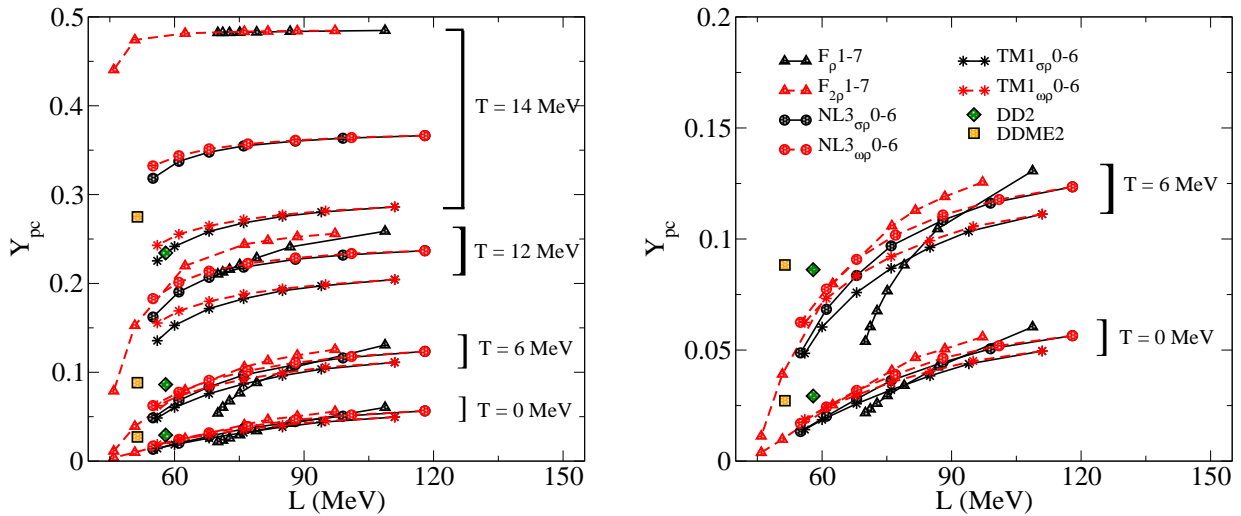}
 \caption{(Color online)  Critical proton fraction, $Y_{pc}$, as a function of $L$
 for several temperatures, and for all the models considered in this study. The right panel shows the results for $T=0$ and 6 MeV only. }
 \label{fig6}
 \end{figure*}

We will start with the  analysis of the effects of temperature on the spinodal sections obtained for the models with
extreme values of $L$, in particular  the largest and the lowest of
each family.
In Figs. \ref{fig4} and \ref{fig5},
we plot the spinodal sections of members 1 and 7 of the $F_\rho$ and
$F_{2\rho}$ families for $T=0,\, 6,\, 12 $ and 14 MeV, and, in Fig.  \ref{fig4}, we also represent  the line of
critical points by a dashed line. At these
points, which are common to both the binodal and the spinodal, the
direction of the instability is parallel to the tangent at the
spinodal, and the pressure is maximum.
 Some conclusions are
in order: a) the behavior with temperature is similar to the one
obtained in \cite{avancini06}, the larger the temperature the smaller
the spinodal section and matter is more symmetric inside the
spinodal. Eventually, at the critical temperature, the section is
reduced to a point and, for larger temperatures, homogeneous matter is
always stable; b) the spinodal sections of models  $F_\rho$1 and
$F_{2\rho}1$, left panels of Fig. \ref{fig4}, are very similar,
as expected, because, these two models have very similar
properties (see also Fig. \ref{fig3}): they are the models with the largest slope $L$ and the strength  of the cross-couplings is very small; c) the
same is not true for the members with the smallest values of $L$, $F_\rho7$ and
$F_{2\rho}7$. The spinodal of the $F_{2\rho}7$ model becomes
larger, extending to larger asymmetries and densities. This same
behavior was obtained with the NL3 and TM1 families, and has been
discussed in \cite{pais2010,paisVlasov}, but for dynamical
spinodals. The $F_\rho$ family shows a different behavior, and the spinodal
of the model with the smallest $L$, $F_\rho7$,  is smaller than
$F_\rho1$. This may be attributed
  to the different behavior of  the symmetry energy for
  this model, as can be seen from Fig. \ref{fig3}.
 In Fig. \ref{fig5}, where we compare the four models at different
temperatures, it is clear that $F_{2\rho}7$
is the one for which the spinodal section extends to a
larger range of densities and asymmetries. This behavior is expected since this is the model
with the smallest $L$.

  \begin{table}
 \caption{
 Critical temperatures, and their correspondent critical densities and pressures for all the
models considered in this work. The proton fraction is 0.5.} 
 \label{tab2}
 \begin{ruledtabular}
 \begin{tabular}{cccc}
 Model  & $T_c$ (MeV) & $\rho_c$ (fm$^{-3}$) & $P_c$ (MeVfm$^{-3}$) \\
 \hline
        DD2 &  13.73 & 0.0452 & 0.1785 \\
        DD-ME2 & 13.12 & 0.0445 & 0.1556 \\
        $F_{x\rho}$ &  14.01 & 0.0444 & 0.1802    \\
   NL3 & 14.55 & 0.0463 & 0.1999 \\
   TM1 & 15.62 & 0.0486 & 0.2365 \\
 \end{tabular}
 \end{ruledtabular}
 \end{table} 
 
 \begin{figure*}[htb]
 \includegraphics[width=1\textwidth]{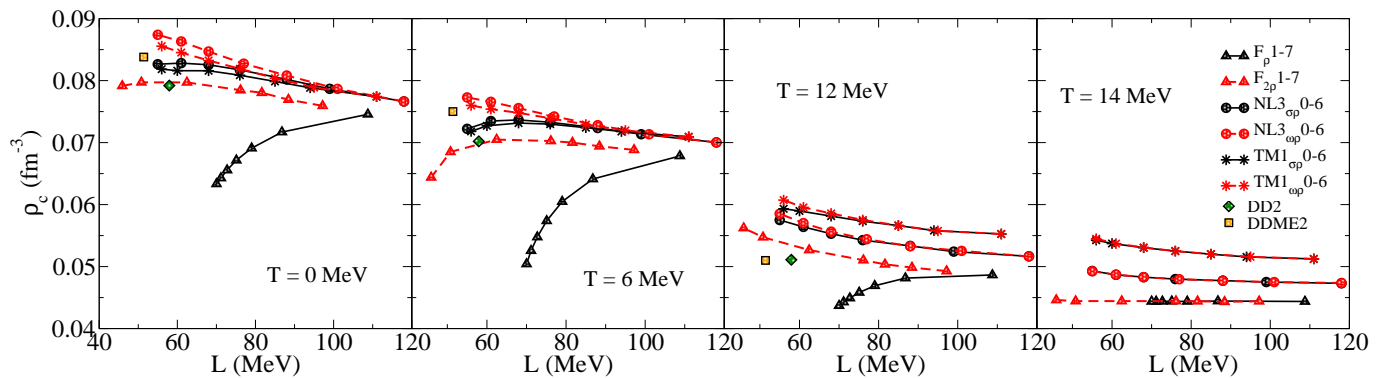} \\
 \caption{(Color online)   Critical density, $\rho_c$, as a function of $L$,
     for several temperatures, and for all the models considered in this study.}
  \label{fig7}
 \end{figure*} 
 
 We now consider the variations of the critical density and 
proton fraction with the temperature and the symmetry energy slope parameter.
 Before embarking on this, we would
like to discuss briefly the results for the critical temperature.

The critical  temperature is totally defined by the
isoscalar properties of the model and, therefore, it is the same
for models that only differ on the isovector properties: the critical temperatures
 for NL3${x\rho}$ ,TM1${x\rho}$ and $F_{x\rho}$ are the same as those for the 
corresponding base models NL3, TM1, BKA22, respectively. 
  The values of the critical  temperature, density and pressure for the base models, 
as well as for the  DD2 and DDME2 models, are given in Table  \ref{tab2}.
  For the  BKA22 model, the critical temperature is very
close to 14 MeV, while for the
TM1 model, the critical temperature is above 15 MeV, and for NL3,
$T_c=14.55$ MeV. 
The TM1 and NL3 $T_c$ values fall inside the interval of temperatures $14.2\le T_c\le
16.1$ MeV, obtained in
\cite{lourenco16}  from a set of 
RMF models with non-linear $\sigma$ terms that have an effective mass at
saturation that reproduces
finite nuclei spin-orbit splittings, and  an
incompressibility in the range 250$\le K_0\le 315$ MeV, as proposed in
\cite{stone14}, and the critical temperature for the BKA22 model lies
very close to the bottom limit. While the incompressibilities for TM1 and
NL3,  281 and 272 MeV, respectively, lie inside the range considered \cite{lourenco16}, for the BKA22 models, it is 220 MeV, and, therefore, it is outside that interval.
However, the critical temperatures
predicted by the models in the present study are far from the value $T_c=17.9\pm 0.4$ MeV
obtained in  \cite{elliott13} from the analysis of six different sets of
experimental data
from heavy-ion reactions.
Let us stress that three
of the models considered (DD2, DDME2 and NL3$\omega\rho6$) went through a set of laboratorial  and
theoretical constraints for
neutron matter, besides predicting star masses above $2\, M_\odot$, as
identified in \cite{fortin16}, and they predict critical temperatures
below 14.55 MeV, even below 14 MeV. Besides these three models, also
models NL3$\sigma\rho6$, TM1$\omega\rho6$,  TM1$\sigma\rho6$ , $F_{2\rho}6$
satisfy most of these constraints: TM1 models
have an incompressibility outside the range considered in
\cite{fortin16}, but well inside the range proposed in \cite{stone14},
and $F_{2\rho}6$ predicts a maximum neutron star mass just below $2M_\odot$. In
\cite{elliott13}, the authors have performed a quite complete
compilation of theoretical predictions  for the critical temperature,
and, in fact,  the RMF models that predict
a critical temperature close to $T_c=17.9\pm 0.4$ MeV do not satisfy
most of the laboratorial constraints at saturation density or below.
Thus, one conclusion that  can be drawn is that
the theoretical critical temperature predicted  by the models fitted to the 
ground state properties of finite nuclei and nuclear matter, and
satisfying the $2M_\odot$ constraint, does not agree with the experimentally
extracted value of the critical temperature.

   \begin{figure*}[htb]
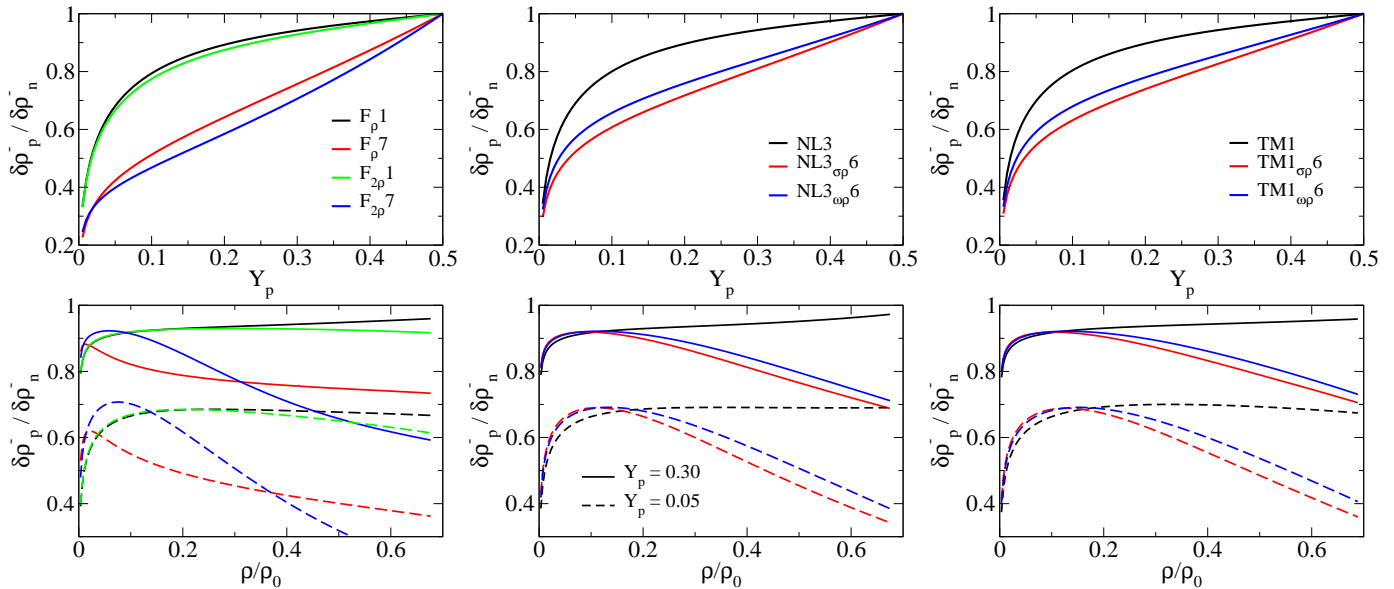

 \centering
\begin{tabular}{ccc}
 \includegraphics[width=0.33\linewidth,angle=0]{fig8a.eps}&
\includegraphics[width=0.33\linewidth,angle=0]{fig8b.eps}&
 \includegraphics[width=0.33\linewidth,angle=0]{fig8c.eps}
\end{tabular}
 \caption{(Color online) The fluctuations $\delta\rho_p^{-}$/$\delta\rho_n^{-}$ at $T = 0 \hbox{ MeV}$ as a function of 
 the  proton fraction $Y_P$ (top panels)  with $\rho = 0.06
 \hbox{ fm}^{-3}$, and  as a function of $\rho/\rho_0$ (bottom panels), with $Y_p=$ 0.30 (solid), and 0.05 (dashed). The calculations shown are for the models $F_\rho$ and
  $F_{2\rho}$ (left), $NL3\omega\rho$ and $NL3\sigma\rho$ (middle) and
  $TM1\omega\rho$ and $TM1\sigma\rho$ (right panels).
}
 \label{fig8}
 \end{figure*}

   \begin{figure*}[htb]
 \centering
 \includegraphics[width=0.95\linewidth,angle=0]{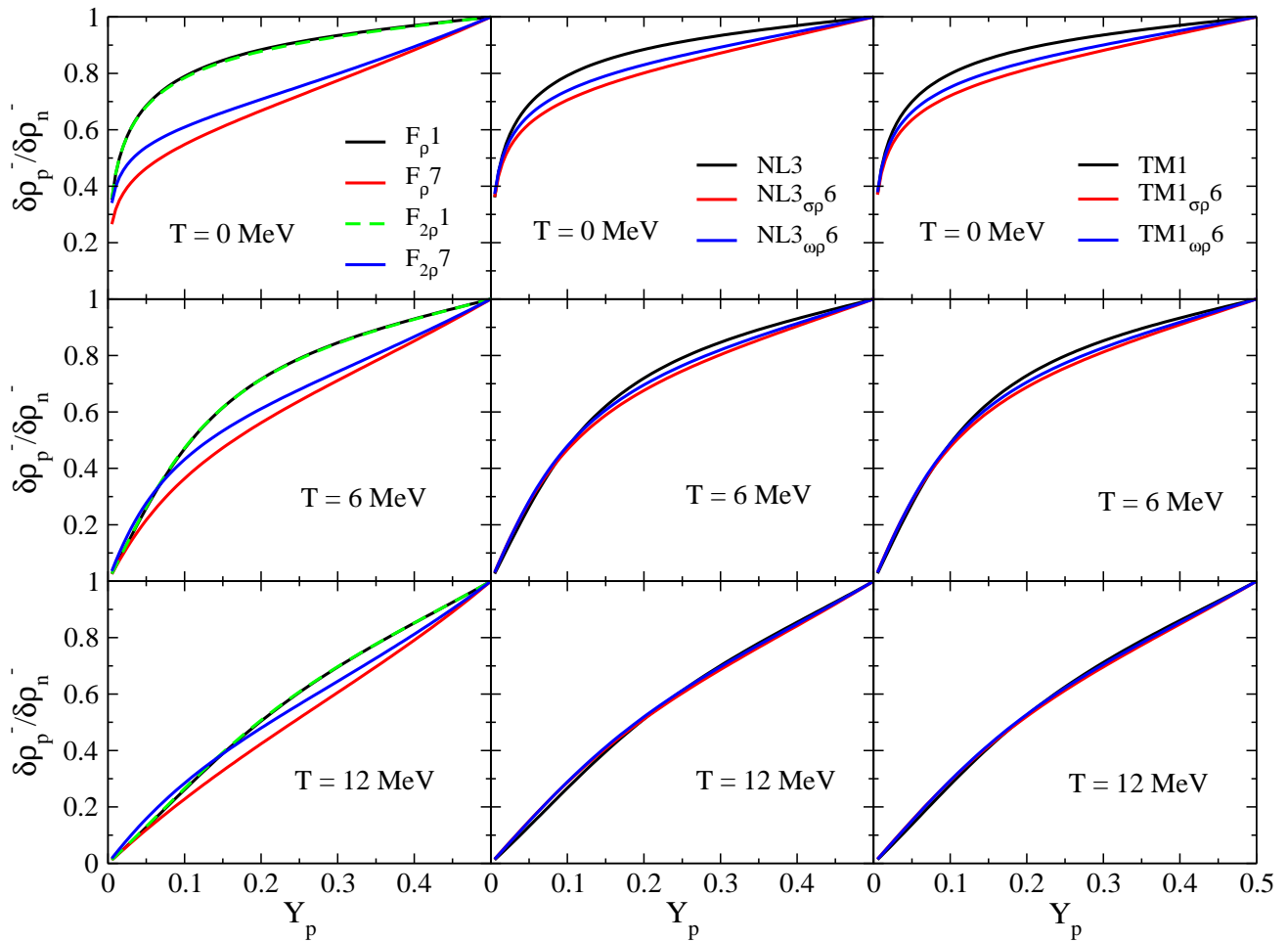} \\ 
 \caption{(Color online) The fluctuations $\delta\rho_p^{-}$/$\delta\rho_n^{-}$ as a function of the proton fraction, $Y_P$, for a fixed baryon density of $\rho = 0.04 \hbox{ fm}^{-3}$ at $T = 0 \hbox{ MeV}$ (top), $T = 6 \hbox{ MeV}$ (middle), and $T = 12 \hbox{ MeV}$ (bottom panels), for the F$_{x\rho}$ (left), NL3$x\rho$ (middle), and TM1$x\rho$ (right) families.}
 \label{fig9}
 \end{figure*}

   \begin{figure*}[htb]
 \centering
 \includegraphics[width=0.95\linewidth,angle=0]{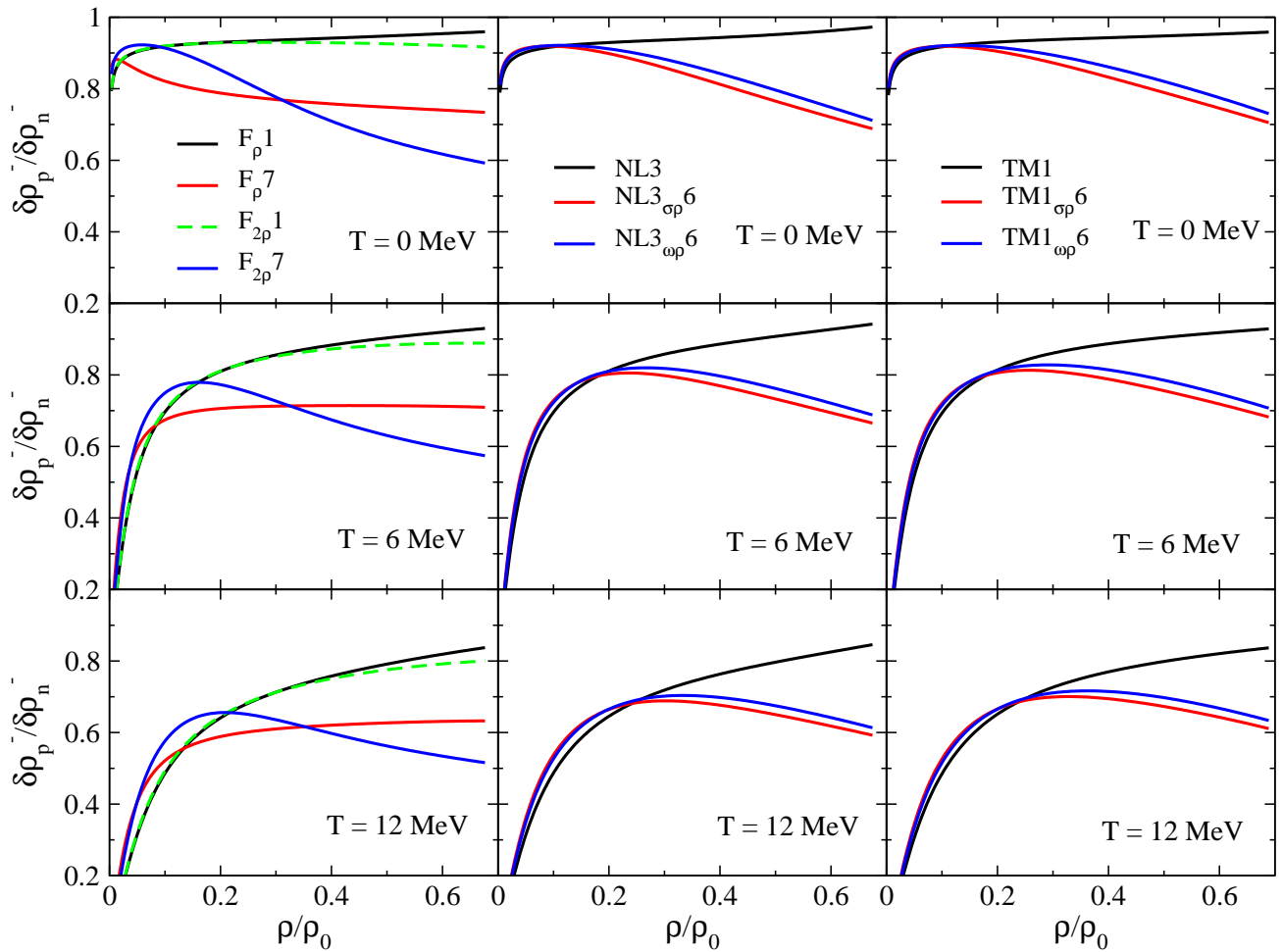} \\
 \caption{(Color online) The fluctuations $\delta\rho_p^{-}$/$\delta\rho_n^{-}$ as a function of $\rho/\rho_0$, for a fixed proton fraction of $Y_p = 0.3$, at $T = 0 \hbox{ MeV}$ (top), $T = 6 \hbox{ MeV}$ (middle), and $T = 12 \hbox{ MeV}$ (bottom panels), for the F$_{x\rho}$ (left), NL3$x\rho$ (middle), and TM1$x\rho$ (right) families.
}
 \label{fig10}
 \end{figure*}

The critical points give us an indication of the phase space region where
non-homogeneous matter is expected. The critical densities for neutron
rich matter and respective proton fraction for $T=0,6,12$ and 14 MeV
are given in Table \ref{tab3}, and displayed as a function of $L$ in
Figs.  \ref{fig6} and \ref{fig7}. The largest temperature
considered, 14 MeV,  is very close to the critical temperature of the $F_\rho$
and $F_{2\rho}$ models.
 Above the critical temperature,  the models
do not present instabilities and the formation of clusters
is not expected.
We first discuss the critical proton fraction. This quantity tells us
that matter at the critical density with smaller proton fractions is stable against
clusterization at the temperature considered. In the left panel of
Fig. \ref{fig6}, the critical proton fractions are given for all the
temperatures considered. This allows us to see the dependence of the critical
proton fraction on the temperature. The right panel of Fig. \ref{fig6}
shows the same but more extensively for $T=0$ and $6$ MeV.
There is a clear dependence of the critical proton fraction on the slope $L$ (see right
panel of Fig. \ref{fig6} for more details).
The critical proton fraction increases when $L$ increases: this
behavior is valid for all the temperatures considered. However, it should be
pointed out that the smaller the $L$, the softer is the increase of the
$Y_{pc}$ with temperature, for temperatures well below the critical
temperature, and this results in a much wider range of critical
protons fractions at finite temperature than at $T=0$.
For instance, the critical proton fractions for the $F_{2\rho}7 (F_\rho1)$ associated with
smaller (larger) values of $L$ are 0.0039 (0.0604), 0.0112 (0.1307), and 0.0788 (0.2587)  at 
temperature T = 0, 6, and 12 MeV, respectively.

We also notice that the spread in the values of the critical proton fraction, at a given $L$,
among the various models considered, increases with temperature. 
For a given $L$, the
spread of values is not larger than $\sim  0.01$ at $T=0$ MeV. At $T=6$ (12) MeV the
critical proton fractions spread over at least $\sim 0.03$ ($\sim 0.1$),
for a fixed $L$. The largest temperature considered is almost
coincident with, or close to,  the critical temperature of the models
under study. It is
striking that there  can be a  difference of $\sim 0.25$ between the
proton fractions of these models.
 Taking as reference $L\sim 56$
MeV, a value within the constraints imposed by experiments,
$Y_{pc}$ varies between 0.018 and 0.023 for $T=0$ MeV,
0.048 and 0.065 for $T=6$ MeV, 0.136 and 0.186 for $T=12$ MeV, and between
0.225 and 0.478 for $T=14$ MeV.
These trends indicate that the models which are calibrated  using bulk ground state properties of
the finite nuclei do not constrain very well the values of the critical proton fractions at finite temperatures.
In fact, it should be pointed out that the large spread on the critical proton fraction close to 14
MeV results from the fact that for some models, BKA22 (base model for $F_\rho$ and $F_{2\rho}$ families), this
temperature is very close to the critical temperature, while for the
TM1 models, the critical temperature is above 15 MeV, and for NL3,
$T_c=14.55$ MeV.
 Temperatures of the order 5 - 12 MeV
occur in core collapse supernova matter.
We may, therefore, expect a different
evolution of the supernova when different models are considered as the
underlying model of the simulation.
In the neutrino trapped phase, a typical proton fraction is
0.3, and we conclude from the left panel of Fig. \ref{fig6}  that while
for NL3, $F_\rho$, and $F_{2\rho}$, matter at $T=14$ MeV is not
clusterized, for TM1, nonhomogeneous matter still occurs under these
conditions. As a reference we also include the critical proton fractions and the critical densities of
the models DD2 and DDME2 in Figs. \ref{fig6} and \ref{fig7}, since these
  models satisfy many  well established properties. They both have a
  critical temperature below 14 MeV. At $T=0$, they show a proton
  fraction above the predicted one by the model with a similar
  symmetry energy of the six families studied. This difference grows as
  the temperature increases, because 
 these models have a lower critical temperature than all the others.

Let us now discuss how the critical density, $\rho_c$, changes with $L$
and $T$. In Fig. \ref{fig7}, the critical densities are plotted for the
different models and temperatures considered. The model $F_\rho$
stands out because it is the only one that presents a critical density
that increases when $L$ increases, for all temperatures. The density
dependence of the symmetry energy in this model is determined by the
term $\sigma\rho^2$, while all the others have a term  $\sigma^2\rho^2$
or $\omega^2\rho^2$.  Models $F_{2\rho}$, NL3$\sigma\rho$ and
TM1$\sigma\rho$ also show this trend for the lowest temperatures
considered, 0 and 6 MeV, and $L\lesssim 60$ MeV. In all other cases,
$\rho_c$ decreases when $L$ increases. The critical densities of
  models DD2 and DDME2 agree  with the other models.
Taking again $L=56$ MeV
as reference, $\rho_c$ decreases with $T$, from 0.080-0.087 fm$^{-3}$ at $T=0$,
to  0.044-0.054 fm$^{-3}$ at $T=14$ MeV, while the spread of $\rho_c$ increases slightly with
temperature, from 0.006 fm$^{-3}$ at $T=0$ MeV to   0.01 fm$^{-3}$ at $T=
14$ MeV. This quantity seems, therefore,  to be more constrained than
the critical proton fraction.

We address next the distillation effect referred in
\cite{chomaz96,avancini06} within the models under
discussion. This is possible from the analysis of the  instability
direction given by $\delta\rho_p^-/\delta\rho_n^-$. This quantity,
calculated at $T=0$, has
been plotted in Fig. \ref{fig8} 
 as a function of the 
 proton fraction, $Y_P$, for $\rho = 0.06
 \hbox{ fm}^{-3}$  in the top panels, and as
 a function of the density divided by the nuclear saturation density, $\rho/\rho_0$, in the bottom panels, and two
  different values of $Y_p$ (0.30 and 0.05) for the models $F_\rho$ and
  $F_{2\rho}$ (left), NL3$\omega\rho$ and NL3$\sigma\rho$ (middle) and
  TM1$\omega\rho$ and TM1$\sigma\rho$ (right panels). The two proton fractions
  considered are of the order of the proton fractions expected in cold
  catalyzed stellar matter. It is seen that the
  distillation effect is present in all models, the direction of
  instability favors a more isospin symmetric dense matter and a more asymmetric
  gas phase. However, there is a clear difference between models with
  a large $L$ and a small $L$: the distillation effect is much
  stronger for the first ones, and for a fixed proton fraction, the
  distillation effect 
  increases with density, while for the second ones, after a maximum
  attained at $\sim 0.02$ fm$^{-3}$, the ratio
  $\delta\rho_p^-/\delta\rho_n^-$ decreases as the density
  increases. A similar behavior was obtained for density dependent
  models in \cite{avancini06}.  While $F_{2\rho}7$ has a behavior
  very similar to NL3$x\rho7$ and TM1$x\rho7$ models, with $x=\sigma$
  or $\omega$,  once more the $F_{\rho}7$ shows a particular behavior,
  showing a smaller (larger) distillation effect for $\rho<(>) 0.04$
  fm$^{-3}$ than the other models with a similar $L$.
Below saturation density, models with a
  smaller $L$ have larger symmetry energies that disfavor a strong
  distillation effect.

In Figs. \ref{fig9} and \ref{fig10}, the quantities
$\delta\rho_p^-/\delta\rho_n^-$ are plotted for different
temperatures.  We have considered the density 0.04 fm$^{-3}$ in the
set of plots of Fig. \ref{fig9} because this is the density that
corresponds to clusterized matter at all temperatures. 
It is evident that the dependence of the distillation effects  on the symmetry
 energy slope parameter gets washed out with the temperature, and for $T=6$ MeV, the differences are already small, although there are still noticeable differences for the $F_{x\rho}$
families.

In Fig. \ref{fig10}, the proton fraction has been fixed to a
typical value that occurs  in trapped neutrino matter, $y_p=0.3$, and
the dependence of $\delta\rho_p^-/\delta\rho_n^-$ on the density is
shown for different temperatures. Models of the TM1 and NL3 families are
different above densities $\rho\sim 0.03-0.04$ fm$^{-3}$, with the
models with smaller slopes $L$ showing a decrease of the ratios, with
a larger effect on the models with a $\sigma^2\rho^2$ non-linear term.
Models of the $F_{x\rho}$ families show larger differences at all temperatures,
with the small $L$ models having larger
$\delta\rho_p^-/\delta\rho_n^-$ values below $\rho\sim 0.02$
fm$^{-3}$. The $F_{\rho}7$ model differs again from all the other
models with a similar $L$, showing a  $\delta\rho_p^-/\delta\rho_n^-$
that increases monotonically with $\rho$ at finite $T$.

  \setlength{\tabcolsep}{5pt}
 \begin{table*}[t]
 \caption{
Critical densities, $\rho_c$, and proton fractions, $Y_{pc}$, for
different temperatures, and for all the models considered.
The slope parameter $L$ and temperature $T$ are in MeV. The critical density $\rho_c$ is in fm$^{-3}$.} \label{tab3}
  \begin{tabular}{cc ccc ccc ccc cc}
    \hline
    \hline
     &  &  \multicolumn{2}{c}{$T=0$} & \phantom{a} &\multicolumn{2}{c}{$T=6$} &\phantom{a} &\multicolumn{2}{c}{$T=12$} &\phantom{a} &\multicolumn{2}{c}{$T=14$} \\
     \hline
Model & $L$  &  \multicolumn{1}{c}{$\rho_c$} & \multicolumn{1}{c}{$Y_{pc}$}  &\phantom{a} &  \multicolumn{1}{c}{$\rho_c$} & \multicolumn{1}{c}{$Y_{pc}$} &\phantom{a} &  \multicolumn{1}{c}{$\rho_c$} & \multicolumn{1}{c}{$Y_{pc}$}  &\phantom{a} &  \multicolumn{1}{c}{$\rho_c$} & \multicolumn{1}{c}{$Y_{pc}$} \\[0.5pt]
    \hline
    DD2 & 57.94 & 0.0792 & 0.0293 & \phantom{a} & 0.0702 & 0.0862 & \phantom{a} & 0.0511 & 0.2343 & \phantom{a} & - & -\\
    DD-ME2 & 51.4 & 0.0838 & 0.0272 & \phantom{a} & 0.0750 & 0.0883 & \phantom{a} & 0.0510 & 0.2749 & \phantom{a} & - & - \\
    \\
 $F_{\rho}$1 &   108.77  &   0.0746  &  0.0604  & \phantom{a} &  0.0678 &   0.1307 &\phantom{a}&   0.0486 &   0.2587 & \phantom{a} &   0.0444 &   0.4848 \\
 $F_{\rho}$2 &    86.77  &   0.0717  &  0.0424  & \phantom{a} &  0.0641 &   0.1046 &\phantom{a}&   0.0481 &   0.2410 & \phantom{a} &   0.0444 &   0.4836 \\   
 $F_{\rho}$3 &    79.02  &   0.0691  &  0.0341  & \phantom{a} &  0.0605 &   0.0883 &\phantom{a}&   0.0469 &   0.2287 & \phantom{a} &   0.0444 &   0.4829 \\
 $F_{\rho}$4 &    75.10  &   0.0672  &  0.0294  & \phantom{a} &  0.0574 &   0.0766 &\phantom{a}&   0.0458 &   0.2210 & \phantom{a} &   0.0444 &   0.4825 \\
 $F_{\rho}$5 &    72.74  &   0.0656  &  0.0260  & \phantom{a} &  0.0548 &   0.0676 &\phantom{a}&   0.0449 &   0.2159 & \phantom{a} &   0.0444 &   0.4824 \\
 $F_{\rho}$6 &    71.16  &   0.0643  &  0.0234  & \phantom{a} &  0.0526 &   0.0605 &\phantom{a}&   0.0443 &   0.2127 & \phantom{a} &   0.0444 &   0.4823 \\
 $F_{\rho}$7 &    70.02  &   0.0633  &  0.0217  & \phantom{a} &  0.0504 &   0.0539 &\phantom{a}&   0.0437 &   0.2105 & \phantom{a} &   0.0444 &   0.4823 \\
 \\
 $F_{2\rho}$1 &    97.19  &   0.0759  &  0.0559  & \phantom{a} &  0.0688 &   0.1256 &\phantom{a}&   0.0493 &   0.2560  &\phantom{a}&  0.0444 &   0.4845 \\
 $F_{2\rho}$2 &    88.44  &   0.0769  &  0.0505  & \phantom{a} &  0.0694 &   0.1191 &\phantom{a}&   0.0498 &   0.2525  &\phantom{a}&  0.0443 &   0.4842 \\
 $F_{2\rho}$3 &    81.62  &   0.0780  &  0.0466  & \phantom{a} &  0.0700 &   0.1129 &\phantom{a}&   0.0504 &   0.2483  &\phantom{a}&  0.0444 &   0.4839 \\
 $F_{2\rho}$4 &    76.17  &   0.0785  &  0.0408  & \phantom{a} &  0.0703 &   0.1059 &\phantom{a}&   0.0510 &   0.2440  &\phantom{a}&  0.0444 &   0.4835 \\
 $F_{2\rho}$5 &    62.45  &   0.0797  &  0.0253  & \phantom{a} &  0.0705 &   0.0799 &\phantom{a}&   0.0527 &   0.2198  &\phantom{a}&  0.0444 &   0.4815 \\
 $F_{2\rho}$6 &    50.80  &   0.0797  &  0.0098  & \phantom{a} &  0.0685 &   0.0391 &\phantom{a}&   0.0547 &   0.1527  &\phantom{a}&  0.0444 &   0.4741 \\
 $F_{2\rho}$7 &    45.91  &   0.0791  &  0.0039  & \phantom{a} &  0.0644 &   0.0112 &\phantom{a}&   0.0562 &   0.0788  &\phantom{a}&  0.0446 &   0.4405 \\
 \\
NL3 &    118.00  &   0.0766 &   0.0565 & \phantom{a} &   0.0700 &   0.1235 &\phantom{a}&   0.0517 &   0.2369 & \phantom{a} &   0.0473 &   0.3664 \\
\\
 NL3${\sigma\rho}$1 &    99.00   &  0.0787  &  0.0506  & \phantom{a} &  0.0713  &  0.1162  &\phantom{a}&  0.0524  &  0.2319  & \phantom{a} &  0.0475  &  0.3634  \\
 NL3${\sigma\rho}$2 &    88.00   &  0.0802  &  0.0445  & \phantom{a} &  0.0724  &  0.1085  &\phantom{a}&  0.0533  &  0.2272  & \phantom{a} &  0.0477  &  0.3602  \\
 NL3${\sigma\rho}$3 &    76.00   &  0.0817  &  0.0363  & \phantom{a} &  0.0733  &  0.0969  &\phantom{a}&  0.0543  &  0.2183  & \phantom{a} &  0.0480  &  0.3548  \\
 NL3${\sigma\rho}$4 &    68.00   &  0.0825  &  0.0279  & \phantom{a} &  0.0737  &  0.0836  &\phantom{a}&  0.0553  &  0.2069  & \phantom{a} &  0.0483  &  0.3478  \\
 NL3${\sigma\rho}$5 &    61.00   &  0.0828  &  0.0202  & \phantom{a} &  0.0735  &  0.0683  &\phantom{a}&  0.0564  &  0.1905  & \phantom{a} &  0.0487  &  0.3375  \\
 NL3${\sigma\rho}$6 &    55.00   &  0.0826  &  0.0133  & \phantom{a} &  0.0722  &  0.0487  &\phantom{a}&  0.0575  &  0.1622  & \phantom{a} &  0.0493  &  0.3184  \\
 \\
 NL3${\omega\rho}$1 &   101.00   &  0.0787  &  0.0519  & \phantom{a} &  0.0713  &  0.1178  &\phantom{a}&  0.0525  &  0.2337  & \phantom{a} &  0.0475  &  0.3642  \\
 NL3${\omega\rho}$2 &    88.00   &  0.0808  &  0.0464  & \phantom{a} &  0.0728  &  0.1108  &\phantom{a}&  0.0533  &  0.2287  & \phantom{a} &  0.0477  &  0.3611  \\
 NL3${\omega\rho}$3 &    77.00   &  0.0827  &  0.0390  & \phantom{a} &  0.0742  &  0.1017  &\phantom{a}&  0.0544  &  0.2226  & \phantom{a} &  0.0480  &  0.3570  \\
 NL3${\omega\rho}$4 &    68.00   &  0.0847  &  0.0318  & \phantom{a} &  0.0756  &  0.0908  &\phantom{a}&  0.0556  &  0.2138  & \phantom{a} &  0.0483  &  0.3515  \\
 NL3${\omega\rho}$5 &    61.00   &  0.0863  &  0.0244  & \phantom{a} &  0.0766  &  0.0775  &\phantom{a}&  0.0570  &  0.2013  & \phantom{a} &  0.0487  &  0.3437  \\
 NL3${\omega\rho}$6 &    55.00   &  0.0874  &  0.0170  & \phantom{a} &  0.0773  &  0.0625  &\phantom{a}&  0.0585  &  0.1829  & \phantom{a} &  0.0492  &  0.3323  \\
 \\
 TM1 &    111.00  &   0.0774 &   0.0496 & \phantom{a} &   0.0709 &   0.1112 &\phantom{a}&   0.0553 &   0.2044 & \phantom{a} &   0.0512 &   0.2862 \\
 \\
 TM1${\sigma\rho}$1 &    94.00   &  0.0788  &  0.0438  & \phantom{a} &  0.0718  &  0.1034  &\phantom{a}&  0.0558  &  0.1972  & \phantom{a} &  0.0516  &  0.2803  \\
 TM1${\sigma\rho}$2 &    85.00   &  0.0799  &  0.0384  & \phantom{a} &  0.0724  &  0.0962  &\phantom{a}&  0.0566  &  0.1917  & \phantom{a} &  0.0520  &  0.2754  \\
 TM1${\sigma\rho}$3 &    76.00   &  0.0809  &  0.0321  & \phantom{a} &  0.0730  &  0.0868  &\phantom{a}&  0.0573  &  0.1828  & \phantom{a} &  0.0525  &  0.2681  \\
 TM1${\sigma\rho}$4 &    68.00   &  0.0815  &  0.0258  & \phantom{a} &  0.0732  &  0.0759  &\phantom{a}&  0.0581  &  0.1717  & \phantom{a} &  0.0530  &  0.2585  \\
 TM1${\sigma\rho}$5 &    60.00   &  0.0818  &  0.0186  & \phantom{a} &  0.0727  &  0.0604  &\phantom{a}&  0.0590  &  0.1528  & \phantom{a} &  0.0537  &  0.2417  \\
 TM1${\sigma\rho}$6 &    56.00   &  0.0817  &  0.0142  & \phantom{a} &  0.0718  &  0.0485  &\phantom{a}&  0.0593  &  0.1356  & \phantom{a} &  0.0543  &  0.2254  \\
 \\
 TM1${\omega\rho}$1 &     95.00  &   0.0789  &  0.0451  & \phantom{a} &  0.0720 &   0.1056 &\phantom{a}&   0.0558 &   0.1988 & \phantom{a} &   0.0516  &  0.2817 \\
 TM1${\omega\rho}$2 &     85.00  &   0.0804  &  0.0404  & \phantom{a} &  0.0729 &   0.0993 &\phantom{a}&   0.0566 &   0.1941 & \phantom{a} &   0.0520  &  0.2773 \\
 TM1${\omega\rho}$3 &     76.00  &   0.0819  &  0.0354  & \phantom{a} &  0.0739 &   0.0921 &\phantom{a}&   0.0575 &   0.1878 & \phantom{a} &   0.0525  &  0.2718 \\
 TM1${\omega\rho}$4 &     68.00  &   0.0832  &  0.0297  & \phantom{a} &  0.0748 &   0.0834 &\phantom{a}&   0.0586 &   0.1799 & \phantom{a} &   0.0531  &  0.2648 \\
 TM1${\omega\rho}$5 &     61.00  &   0.0845  &  0.0243  & \phantom{a} &  0.0754 &   0.0732 &\phantom{a}&   0.0595 &   0.1691 & \phantom{a} &   0.0537  &  0.2554 \\
 TM1${\omega\rho}$6 &     56.00  &   0.0856  &  0.0189  & \phantom{a} &  0.0760 &   0.0624 &\phantom{a}&   0.0607 &   0.1555 & \phantom{a} &   0.0545  &  0.2430 \\
 \hline
\end{tabular}  
\end{table*}

\section{Conclusion}

In the present study, we have analysed the extension of the nonhomogeneous
nuclear matter in the  density, isospin and temperature directions,
as predicted by six different families of the RMF models, together
with two density-dependent models.  The six families of models have
been built  from three different base models, whose parameters are
fitted to the ground state properties of nuclei. An extra term that
couples the $\rho$-meson either to the $\sigma$  or $\omega$-meson is
appropriately added to each of the base models to yield the variation
in the symmetry energy slope $L$ approximately between 50 and 100 MeV
\cite{Horowitz,Carriere,frho}.  The thermodynamical spinodal sections
are determined by the loci in phase space where the curvature matrix of
the free energy is zero. These spinodal sections and lines of critical
points are obtained for temperatures below the critical temperature
above which there is a smooth transition from a gas to a liquid phase.
The critical proton fractions and densities for a given temperature
give us an indication whether clusterized matter could occur under
some particular conditions. In particular, the  clusterized matter is
not expected at densities larger and proton fractions smaller than the
corresponding critical values.

It is shown that for a given symmetry energy slope parameter $L$,
the models that include a non-linear $\sigma-\rho$ cross-coupling
predict smaller critical densities and proton fractions. The effect is
specially strong for the $F_\rho$ family, which includes a $\sigma\rho^2$
cross-coupling term.  The critical density is more constrained. In fact,
considering a slope $L=56$ MeV,   the spread on the critical density
increases from 0.006 fm$^{-3}$ at $T=0$ to $\sim 0.01$ fm$^{-3}$  at
$T=14$  MeV.  The critical proton fraction at zero temperature increases
when the slope $L$ increases, and for a given value $L$, it is almost
independent of the model considered. This is not the case at finite
temperature, where a spread on the proton fraction of 0.25 for $T=14$
MeV is found, when all the different  models are considered.
This large spread on the critical proton fraction close to 14
MeV can be attributed to the different critical temperatures of the
models under study.   Since the models considered predict different
critical temperatures associated with symmetric matter,  the critical
proton fractions at temperatures above 10 MeV may show a large spread.

We have  also analysed the behavior of the distillation effect with
temperature. In particular, previous results, concerning a smaller effect
within models with a smaller slope $L$, were confirmed.  Although the
temperature washes out some of the differences between the models,
mainly among the models of the same family, some differences remain,
the stronger ones among models belonging to the $F_{x\rho}$ families.

 It is observed that the $F_\rho$ family, which includes a cubic
cross-coupling term of the type $\sigma\rho^2$, behaves differently as
compared to the other families of models in which quartic cross-coupling
terms of the type $\sigma^2\rho^2$ or $\omega^2\rho^2$ are considered.
 Five of the six  the families
  contain at least one model that  satisfies the constrains coming from microscopic
calculations for pure neutron matter at sub-saturation densities
(see Fig. \ref{fig1} and Ref. \cite{paisVlasov}),  the  $F_\rho$
family being
the only one that does not satisfy this constrain.

 Seemingly, these results favour
the inclusion of quartic order cross-coupling terms over the cubic order
term, though a cubic term should also be included from ``naturalness''
arguments \cite{mueller96}.  Therefore, a more careful calibration
should be undertaken, which takes into account constrains from nuclear
ground state properties, as well as  constraints coming from microscopic
calculations for neutron matter.

\section*{ACKNOWLEDGMENTS}

Partial support comes from Funda\c c\~ao para a Ci\^encia e Tecnologia (FCT), Portugal,
under the project No. UID/FIS/04564/2016 and from “NewCompStar”, COST Action MP130. 
H.P.  is  supported  by  FCT (Portugal)  under  Project  No. SFRH/BPD/95566/2013. 
N.A. would like to thankfully acknowledge the useful discussions and the nice hospitality extended
to him during his visit to University of Coimbra, when this work started.


\vspace*{1.0 in}
\begin{thebibliography}{99}

\bibitem{CCSN} A. Mezzacappa,  Annu. Rev. Nucl. Part.  Sci. {\bf 55}, 467 (2005); H.-T. Janka, K.~Langanke, A.~Marek, G.~Martinez-Pinedo, and B.~Mueller, Phys. Rept.{\bf 442}, 38  (2007); H.-T. Janka, Ann.  Rev. Nucl. Part. Sci. {\bf 62}, 407  (2012); A.~Burrows, Rev. Mod. Phys. {\bf 85}, 245  (2013).

\bibitem{Stone07} J. R. Stone, and P.-G. Reinhard, Prog. Part. Nucl. Phys. {\bf 58}, 587 (2007). 

\bibitem{Mezzacappa05} A. Mezzacappa,  Annu. Rev. Nucl. Part.  Sci. {\bf 55}, 467 (2005).

\bibitem{oertel16} M. Oertel, M. Hempel, T. Kl\"ahn, and S. Typel, arXiv:1610.03361.

\bibitem{clusters} S. Typel, G. R\"opke, T. Kl\"ahn, D. Blaschke, and H. H.
Wolter, Phys. Rev. C {\bf 81}, 015803 (2010); C. J. Horowitz and A. Schwenk, Nuclear Physics A {\bf 776}, 55 (2006); S. Heckel, P. P. Schneider, and A. Sedrakian, Phys. Rev. C {\bf 80}, 015805 (2009); M. Ferreira and C. Provid\^encia, Phys. Rev. C {\bf 85}, 055811 (2012); S. S. Avancini, C. C. Barros Jr., D. P. Menezes, and C. Provid\^encia, Phys. Rev. C {\bf 82}, 025808 (2010); S. S. Avancini, C. C. Barros Jr., L. Brito, S. Chiacchiera, D. P. Menezes, and C. Provid\^encia, Phys. Rev. C {\bf 85}, 035806 (2012).

\bibitem{pasta} D. G. Ravenhall, C. J. Pethick, and J. R. Wilson, Phys. Rev. Lett. {\bf 50}, 2066 (1983); C. J. Horowitz, M. A. P\'erez-Garc\'ia, and J. Piekarewicz, Phys. Rev. C {\bf 69}, 045804 (2004);  T. Maruyama, T. Tatsumi, D. N. Voskresensky, T. Tanigawa, and S. Chiba, Phys. Rev. C {\bf 72}, 015802 (2005); G. Watanabe, T. Maruyama, K. Sato, K. Yasuoka, and T. Ebisuzaki, Phys. Rev. Lett. {\bf 94}, 031101 (2005); H. Pais and J. R. Stone, Phys. Rev. Lett. {\bf 109}, 151101 (2012); F. Grill, H. Pais, C. Provid\^encia, I. Vida\~na, and S. S. Avancini, Phys. Rev. C {\bf 90}, 045803 (2014); H. Pais, S. Chiacchiera, and C. Provid\^encia, Phys. Rev. C {\bf 91}, 055801 (2015).

\bibitem{coolingPNS} P.~Haensel, Acta Phys.Polon. {\bf B25}, 373 (1994); D.~Page and S.~Reddy, Phys. Rev. Lett. {\bf 111}, 241102  (2013).

\bibitem{shock} R.~D. Williams and S.~E. Koonin, Nucl. Phys. A {\bf 435}, 844  (1985); S.~Furusawa, H.~Nagakura, K.~Sumiyoshi, and S.~Yamada, Astrophys. J. {\bf 774}, 78  (2013).

\bibitem{Buyukcizmeci13} N. Buyukcizmeci, A. Botvina, I. Mishustin, R. Ogul, M. Hempel, J. Schaffner-Bielich, F.-K. Thielemann, S. Furusawa, K. Sumiyoshi, S. Yamada, and H. Suzuki, Nucl.Phys. A {\bf 907}, 13 (2013).

\bibitem{lourenco16} O. Louren\c co, B. M. Santos, M. Dutra, and A. Delfino,  Phys. Rev. C {\bf 94}, 045207 (2016).

\bibitem{experimental} V. A. Karnaukhov, Phys. At. Nucl. {\bf 60}, 1625 (1997); Nucl. Phys. A {\bf 734}, 520 (2004); V. A. Karnaukhov et al., Nucl. Phys. A {\bf 780}, 91 (2006); V. A. Karnaukhov, Phys. At. Nucl. {\bf 71}, 2067 (2008).

\bibitem{natowitz02}  J. B. Natowitz, K. Hagel, Y. Ma, M. Murray, L. Qin, R. Wada, and J. Wang, Phys. Rev. Lett. {\bf 89}, 212701 (2002).

\bibitem{karnaukhov03}  V. A. Karnaukhov, et al., Phys. Rev. C {\bf 67}, 011601(R) (2003).

\bibitem{elliott13}  J. B. Elliott, P. T. Lake, L. G. Moretto, and L. Phair, Phys. Rev. C {\bf 87}, 054622 (2013).

\bibitem{stone14} J. R. Stone, N. J. Stone, and S. A. Moszkowski, Phys. Rev. C {\bf 89}, 044316 (2014).


\bibitem{avanciniPasta} S. S. Avancini, S. Chiacchiera, D. P. Menezes, and C. Provid\^encia, Phys. Rev. C {\bf 82}, 055807 (2010); Phys. Rev. C {\bf 85}, 059904(E) (2012). 

\bibitem{avancini06} S. S. Avancini, L. Brito, Ph. Chomaz, D. P. Menezes, and C. Provid\^encia, Phys. Rev. C {\bf 74}, 024317 (2006). 


\bibitem{pais16} H. Pais, A. Sulaksono, B. K. Agrawal, and C. Provid\^encia, Phys. Rev. C {\bf 93}, 045802 (2016). 

\bibitem{Ducoin11} C. Ducoin, J. Margueron, C. Provid\^encia, and I. Vida\~na, Phys. Rev. C {\bf 83}, 045810 (2011).

\bibitem{quarticA} R. J. Furnstahl, B. D. Serot, and H. B. Tang, Nucl. Phys. A {\bf 598}, 539 (1996).

\bibitem{quarticB} R. J. Furnstahl, B. D. Serot, and H. B. Tang, Nucl. Phys. A {\bf 615}, 441 (1997).

\bibitem{quartic1} A. Sulaksono and Kasmudin, Phys. Rev. C {\bf 80}, 054317 (2009).

\bibitem{quartic2} B. K. Agrawal, Phys. Rev. C {\bf 81}, 034323 (2010).

\bibitem{bctl98} V. Baran, M. Colonna, M. Di Toro, and A. B. Larionov, Nucl. Phys. A {\bf 632}, 287 (1998).

\bibitem{ms} H. M\"uller and B. D. Serot, Phys. Rev. C {\bf 52}, 2072 (1995).

\bibitem{reid} M. Modell and R. C. Reid, Thermodynamics and Its Applications, 2nd edition (Prentice-Hall, Englewood Cliffs, NJ, 1983).

\bibitem{Carriere} J. Carriere, C. J. Horowitz, and J. Piekarewicz, Astrophys. J. {\bf 593}, 463 (2003).

\bibitem{paisVlasov} H. Pais, C. Provid\^encia, Phys. Rev. C {\bf 94}, 015808 (2016).

\bibitem{frho} N. Alam, A. Sulaksono, and B. K. Agrawal, Phys. Rev. C {\bf 92}, 015804 (2015).

\bibitem{bka22} B. K. Agrawal, Phys. Rev. C {\bf 81}, 034323 (2010).

\bibitem{nl3} G. A. Lalazissis, J. K\"onig, and P. Ring, Phys. Rev. C {\bf 55}, 540 (1997).

\bibitem{tm1} Y. Sugahara and H. Toki, Nucl. Phys. A {\bf 579}, 557 (1994);  K. Sumiyoshi, H. Kuwabara, and H. Toki, Nucl. Phys. A {\bf 581}, 725 (1995).

\bibitem{Horowitz} C. J. Horowitz and J. Piekarewicz, Phys. Rev. Lett. {\bf 86}, 5647 (2001). 

\bibitem{J1614} P. B. Demorest, T. Pennucci, S. M. Ransom, M. S. E. Roberts, and J. W. T. Hessels, Nature (London) {\bf 467}, 1081 (2010); E. Fonseca, T. T. Pennucci, J. A. Ellis, I. H. Stairs, D. J. Nice, S. M. Ransom, P. B. Demorest, Z. Arzoumanian, K. Crowter, T. Dolch, R. D. Ferdman, M. E. Gonzalez, G. Jones, M. L. Jones, M. T. Lam, L. Levin, M. A. McLaughlin, K. Stovall, J. K. Swiggum, and W. Zhu, arXiv:1603.00545.

\bibitem{J0348} J. Antoniadis, P. C. C. Freire, N. Wex, T. M. Tauris, R. S. Lynch,
M. H. van Kerkwijk, M. Kramer, C. Bassa, V. S. Dhillon, T.
Driebe, J. W. T. Hessels, V. M. Kaspi, V. I. Kondratiev, N.
Langer, T. R. Marsh, M. A. McLaughlin, T. T. Pennucci, S. M.
Ransom, I. H. Stairs, J. van Leeuwen, J. P. W. Verbiest, and D. G. Whelan, Science {\bf 340}, 448 (2013).

\bibitem{Alam16} N. Alam, B. K. Agrawal, M. Fortin, H. Pais, C. Provid\^encia, Ad. R. Raduta, and A. Sulaksono, Phys. Rev. C {\bf 94}, 052801(R) (2016).

\bibitem{dd2} S. Typel, G. R\"opke, T. Kl\"ahn, D. Blaschke, and H. H. Wolter, Phys. Rev. C {\bf 81}, 015803 (2010).

\bibitem{ddme2} G. A. Lalazissis, T. Niksi\'c, D. Vretenar, and P. Ring, Phys. Rev. C {\bf 71}, 024312 (2005).

\bibitem{fortin16} M. Fortin, C. Provid\^encia, Ad. R. Raduta, F. Gulminelli, J. L. Zdunik, P. Haensel, and M. Bejger, Phys. Rev. C {\bf 94}, 035804 (2016).

\bibitem{Hebeler} K. Hebeler, J. M. Lattimer, C. J. Pethick, and A. Schwenk, Astrophys. J. {\bf 773}, 11 (2013).

\bibitem{Gandolfi} S. Gandolfi, J. Carlson, and S. Reddy, Phys. Rev. C {\bf 85}, 032801 (2012).

\bibitem{flow} P. Danielewicz, R. Lacey, and W. G. Lynch, Science {\bf 298}, 1592 (2002).

\bibitem{kaons} W. G. Lynch, M. B. Tsang, Y. Zhang, P. Danielewicz, M. Famiano, Z. Li, and A. W. Steiner, Prog. Part. Nucl. Phys. {\bf 62}, 427 (2009); C. Fuchs, Prog. Part. Nucl. Phys. {\bf 56}, 1 (2006).

\bibitem{pais2010} H. Pais, A. Santos, L. Brito, C. Provid\^encia, Phys. Rev. C {\bf 82}, 025801 (2010). 

\bibitem{chomaz96} B.  Jacquot,  S.  Ayik,  Ph.  Chomaz,  and M. Colonna, Phys. Lett. B {\bf 383}, 247 (1996).

\bibitem{mueller96} H. M\"uller, B. D. Serot, Nucl.Phys. A {\bf 606}, 508 (1996);
R. J. Furnstahl, B. D. Serot, and H.-B. Tang, Nucl. Phys. A {\bf 615}, 441 (1997); Nucl. Phys. A {\bf 640}, 505 (E) (1998).


\end{thebibliography}
\end{document}